\RequirePackage{fix-cm}
\documentclass[twocolumn]{svjour3}          
\smartqed  
\usepackage[latin1]{inputenc}
\usepackage{ae}
\usepackage{mathtools}
\usepackage{amsfonts}
\usepackage{graphicx}
\usepackage{epstopdf}
\usepackage{amsmath}
\usepackage{longtable}
\usepackage{multirow}
\usepackage{xcolor,colortbl}
\usepackage[normalem]{ulem}

\journalname{Journal of the Brazilian Society of Mechanical Sciences and Engineering}
\begin{document}
	
	\title{Hybrid Model-Based and Data-Driven Wind Velocity Estimator for an Autonomous Robotic Airship
	}
	
	
	\author{	Apolo~Silva~Marton \and
		Andr\'e~Ricardo~Fioravanti
		\and Jos\'e Raul Azinheira
		\and Ely Carneiro de Paiva
	}
	
	
	\institute{A. S. Marton \at
		Department of Computational Mechanics,  School of  Mechanical Engineering, Mendeleyev Street, 200, 13083-180, University of Campinas, Brazil\\
		\email{apolosm@fem.unicamp.br}           
		\and
		A. R. Fioravanti \at
		Department of Computational Mechanics, School of Mechanical Engineering, Mendeleyev Street, 200, 13083-180, University of Campinas, Brazil \\
		\email{fioravanti@fem.unicamp.br}     
		\and
		J.~R.~Azinheira \at
		Department of Mechanical Engineering, Instituto Superior T\'ecnico, Lisbon, 1049-001, Portugal\\
		\email{jose.raul.azinheira@tecnico.ulisboa.pt} 
		\and
		E.~C.~de Paiva \at
		Department of Integrated Systems, School of  Mechanical Engineering, Mendeleyev Street, 200, 13083-180, University of Campinas, Brazil \\
		\email{elypaiva@fem.unicamp.br} 	
	}
	\date{Received: date / Accepted: date}

	\maketitle
	
	\begin{abstract}
		In the context of autonomous airships, several works in control and guidance use wind velocity to design a control law. However, in general, this information is not directly measured in robotic airships. This paper presents three alternative versions for estimation of wind velocity. Firstly, an Extended Kalman Filter is designed as a model-based approach. Then a Neural Network is designed as a data-driven approach. Finally, a hybrid estimator is proposed by performing a fusion between the previous designed estimators: model-based and data-driven. All approaches consider only Global Positioning System (GPS), Inertial Measurement Unit (IMU) and a one dimensional Pitot tube as available sensors. Simulations in a realistic nonlinear model of the airship suggest that the cooperation between these two techniques increases the estimation performance.
	\end{abstract}
	
	\section{Introduction}
	\label{sec:intro}
	Recently, Unmanned Aerial Vehicles (UAVs) become useful in several applications due to their economic efficiency and mobility. For outdoor applications, air related information such as: angle of attack, sideslip and wind velocity, helps to improve the control performance. 
	
	Outdoor airships commonly have a guidance control to track a trajectory. The first idea is that the attitude reference shall be coincident with the reference trajectory attitude. However, there are two situations when this is not desirable: in the presence of wind disturbances (an almost certainty when flying outdoors) and if the objective is ground-hover (since the desired attitude is arbitrarily defined). 	

	Airships, generally have tail surfaces to produce moments, however they do not have an actuator to directly produce a lateral force. Thus, when the airship is to keep a stationary position with respect to a ground target, it must align against the wind and use its forward propulsion to balance the aerodynamic drag. Therefore, by knowing the wind speed 
	we may improve the control performance.For example, in \cite{zhou2019} a path following control approach is presented. The control law designed assume that attack and sideslip angles are measured. Similarly, in \cite{wangYueying2019,zheng2017} the same information is necessary for the path following approaches proposed.

	This work is placed in the context of project DRONI \cite{marton2019}. The project aims to develop an autonomous airship for performing environmental monitoring tasks in remote amazon rain forest areas. Such tasks include aerodynamic flights and ground-hover (i. e. keep a stationary position with respect to a ground target). Therefore, whenever there is wind, the DRONI airship must try to align itself with the relative airspeed, thus reducing the sideslip angle. This implies that guidance control depends on information about wind velocity and attitude. However, measuring such elements is not a trivial task. 
	
	The most common solution is to estimate the wind velocity in order to extract the necessary information about the vehicle motion. The Model-based techniques are the most popular strategies. As an example, in \cite{perry2008}, it is proposed an approach for estimate the angle of attack and sideslip angle by the kinematic equations of motion of an aerobatic UAV. Meanwhile, with the same kinematic equations, in \cite{cho2011} an Extended Kalman Filter (EKF) is proposed for estimating the wind heading and velocity using an aircraft with a single GPS and Pitot tube. In \cite{johansen2015} a wind velocity observer also based in the kinematics is proposed for small UAVs with experimental results. Similarly, in \cite{shen2015} is also proposed an EKF for wind velocity estimation, however applied to a Stationary stratospheric airship in simulation environment. Then in \cite{rhudy2017} are presented four Model-based solutions considering an aircraft with four different possible configurations of sensors.
	
	Recently, the Machine Learning approach has become popular in the field of robotics \cite{ribeiro2019}. The impressive growing of computational resources and increasing acquired data over the years have increased the potential of these Data-driven techniques. These strategies were already introduced in applications such as control of aircraft \cite{chaturvedi2002} and air data estimation for a Micro-UAV \cite{samy2010}. The wind estimation problem is addressed in \cite{allison2019} using a quadrotor and with a Machine-learning approach. However, a Data-driven online estimation of wind velocity for robotic airships is still a challenge.
	
	This paper presents an alternative version of a Model-based wind velocity estimator using the EKF technique similar to the solution presented in \cite{cho2011}, however taking the DRONI airship as a case study. Then, a Data-driven approach of estimation using a Neural Network (NN) is proposed. Finally, a hybrid version that uses both Model-based and Data driven techniques is considered. 	The main tool to validate the proposed estimation approaches is a dynamical realistic non-linear model of an airship in Simulink/MATLAB. This tool is a result of the research group efforts since the project AURORA \cite{elfes2002} which was improved during the projects DIVA \cite{moutinho2016} and DRONI \cite{marton2019}.
	
	
	This paper is organized as follows: the airship nonlinear modeling is summarized in Sect. \ref{sec:airship_model}; then the kinematic equations of motion are analyzed in Sect. \ref{sec:equations_motion}; an EKF is designed for wind velocity estimation in Sect. \ref{sec:ekf_design}; the NN approach of wind velocity estimation is presented in Sect. \ref{sec:nn_design}; the hybrid version of wind velocity estimation is presented in Sect. \ref{sec:hybrid_design}; validating simulations take place in Sect. \ref{sec:results} by establishing a comparison between the proposed approaches and the approach presented by \cite{cho2011}; finally some conclusions are drawn in Sect. \ref{sec:conclusion}.
	
	\section{Airship modeling}
	\label{sec:airship_model}
	
\begin{figure}[htpb]
		\centering
		\includegraphics[angle=0, scale=0.15]{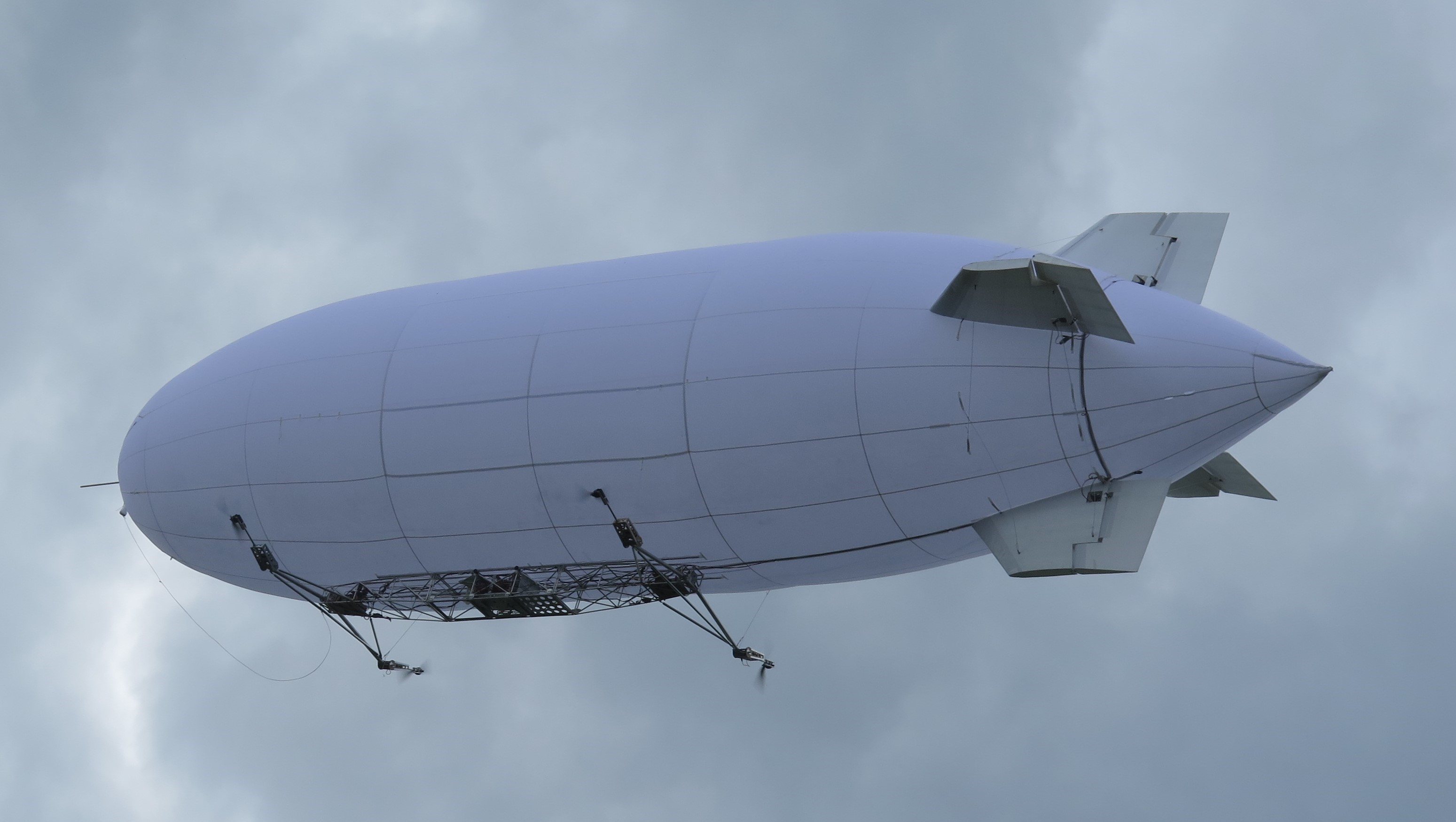}
		\caption{Robotic airship performing its first flight.}
		\label{fig:droni_first_flight}
	\end{figure}

	The DRONI airship has a nonrigid body filled with helium gas, however without a ballonet. It is composed by a hull with 11m length and 2.48m diameter equipped with: 4 vectored propellers with independent thrusters (see Figure \ref{fig:droni_first_flight}) and tail surfaces (rudder, elevator and aileron). 
	
\begin{figure}[htb]
		\centering
		\includegraphics[scale=0.8]{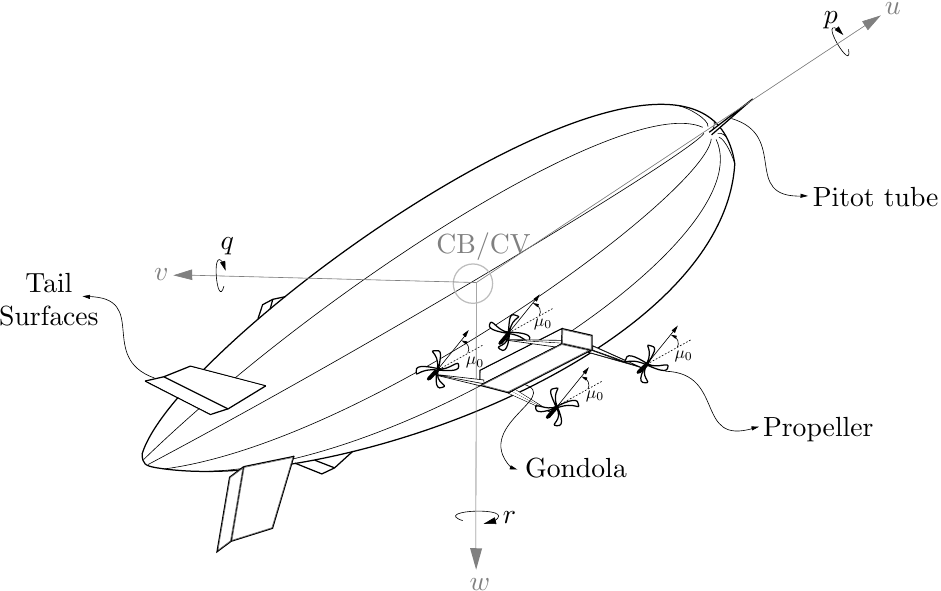}
		\caption{DRONI airship body diagram.}
		\label{fig:airship_cb_cv}
	\end{figure}

		The DRONI airship is instrumented with a Xsens Mti-G 700 which contains an Inertial Measurement Unit (IMU) and a Global Positioning System (GPS). This set of sensors is located at the gondola which is 1.43 meters below the balloon Center of Volume (CV) as shown in Figure \ref{fig:airship_cb_cv}. Additionally,  it is equipped with a one dimensional Pitot tube located at the airship nose. The airship has a total mass of 46.9 kg and buoyancy mass of -43.8 kg, resulting in an apparent mass of 3.1 kg.

	This section presents a summary of the mathematical modeling of an airship.	For a detailed description, please refer to \cite{moutinho2016}. The airship nonlinear model can be expressed as a state-space model given by:

	\begin{subequations}
		\begin{equation}
		\label{eq:kinema_eq}
		\dot{\boldsymbol{\xi}}= g(\boldsymbol{\xi}, \textbf{x})\mathrm{,}
		\end{equation}
		\begin{equation}
		\label{eq:dyn_eq}
		\dot{\textbf{x}}= f(\textbf{x}, \textbf{u}, \textbf{d})\mathrm{,}
		\end{equation}
	\end{subequations}
	where:
	\begin{itemize}
		\item  the kinematic states $\boldsymbol{\xi} = [\textbf{P}_{NED}^T\;\;\boldsymbol{\Phi}^T]^T$ include the cartesian positions $\textbf{P}_{NED} =[ P_N\;\; P_E\;\; P_D]^T$ and angular position $\Phi=[\phi\;\; \theta\;\; \psi]^T$ in the North-East-Down frame;
		\item the dynamic states $\textbf{x} = [\textbf{V}_g^T\;\;\boldsymbol{\Omega}^T]^T$ include the linear speed $\textbf{V}_g=[u\;\; v\;\; w]^T$ and angular speed $\boldsymbol{\Omega}=[ p\;\; q\;\; r]^T$ in the body frame;
		\item the input vector $\textbf{u} = [\delta_e~\delta_a~\delta_r~\delta_0~\mu_0]^T$ includes: $\delta_e, \delta_a$ and $\delta_r$ which are elevator, aileron and rudder deflection (rad); $\delta_0$ as the normalized thrusters voltage (V/V); $\mu_0$ as the common vectoring angle of the thrusters (rad);
		\item and, finally, the disturbance vector $\textbf{d}$ that includes wind velocities and gust parameters.
	\end{itemize}
	
		Another parameter of concern is known as added mass. As stated in \cite{carichner2013book}, the added mass phenomenon is not significant enough to impact airplane performance. Thus, in general, it is not included in the modeling of fixed-wing aircrafts. However, the added mass is always present for any body which is moving through a fluid (see \cite{lamb1918,thomasson2000}). Basically, a body in a fluid behaves as though it has more mass than it actually does. Such behavior varies with the motion nature (accelerating, decelerating or turning). Thus, a significant effect on the dynamics may be experienced by vehicles which have a density similar to the displaced external fluid (such as airships).

		As stated by \cite{moutinho2016,fossen2011}, when the displaced fluid mass is not negligible, as is the case for airships, balloons and submarines, the equations of motion are usually derived from the Lagrangian approach. However, the resultant dynamics can be represented by the Newton-Euler equations including five components of forces and moments, namely: $\textbf{F}_k$ containing the kinetic forces and moments, including Coriolis and centrifugal force terms; $\textbf{F}_a$ given by aerodynamic forces and moments; $\textbf{F}_p$  given by propulsion forces and moments; $\textbf{F}_g$ given by gravity forces and moments, which are function of the difference between the weight and buoyancy forces; and $\textbf{F}_w$ given by the wind induced forces and moments. Therefore, the linear and angular accelerations are given by:
	\begin{equation}
	\label{eq:nonlinear_dynamic}
	f(\textbf{x},\textbf{u},\textbf{d}) = \textbf{M}^{-1}\Big(\textbf{F}_k + \textbf{F}_a + \textbf{F}_p +\textbf{F}_g + \textbf{F}_w\Big)\mathrm{,}
	\end{equation}
	where $\textbf{M}$ includes apparent mass and inertial coefficients of the airship. These equations are referenced in the body frame centered in the Center of Buoyancy (CB) that is approximately equivalent to the Center of Volume (CV) as shown in Figure \ref{fig:airship_cb_cv}. 
	
	Finally, the linear and angular positions are updated through kinematic equations \eqref{eq:kinema_eq2}.
		\begin{equation}
		\label{eq:kinema_eq2}
		g(\boldsymbol{\xi},\textbf{x})= \begin{bmatrix}
		\textbf{S}_\Phi^T & 0_{3\times 3}\\
		0_{3\times 3} & \textbf{R}_\Phi^T
		\end{bmatrix} \begin{bmatrix}
		\textbf{V}_g\\
		\boldsymbol{\Omega}
		\end{bmatrix},
		\end{equation}
		where $0_{3\times 3}$ is a $3\times 3$ null matrix, and $\textbf{S}_\Phi\in \mathbb{R}^{3\times3}$ and $\textbf{R}_\Phi\in \mathbb{R}^{3\times3}$ are rotational matrices from airship body to NED frame, given by:
		\begin{equation*}
		\textbf{S}_\Phi = \begin{bmatrix}\text{c}_\psi \text{c}_\theta & \text{s}_\psi \text{c}_\theta & -s_\theta\\
		\text{c}_\psi \text{s}_\theta \text{s}_\phi - \text{s}_\psi \text{c}_\phi & \text{s}_\psi \text{s}_\theta \text{s}_\phi + \text{c}_\psi \text{c}_\phi & \text{c}_\theta \text{s}_\phi\\
		\text{c}_\psi \text{s}_\theta \text{c}_\phi + \text{s}_\psi \text{s}_\phi & \text{s}_\psi \text{s}_\theta \text{c}_\phi - \text{c}_\psi \text{s}_\phi & \text{c}_\theta \text{c}_\phi\end{bmatrix},
		\end{equation*}
		\begin{equation*}
		\textbf{R}_\Phi = \begin{bmatrix}1 & \text{s}_\phi \text{t}_\theta & \text{c}_\phi \text{t}_\theta\\
		0 & \text{c}_\phi & -\text{s}_\phi\\
		0 & \text{s}_\phi/\text{c}_\theta & \text{c}_\phi/\text{c}_\theta\end{bmatrix},
		\end{equation*}
		where $\text{c}_x = \cos(x)$, $\text{s}_x=\sin(x)$ and $\text{t}_x=\tan(x)$.
	
	\section{Kinematic Equations of motion}
	\label{sec:equations_motion}
	
	In this section we find three equations which are correlated with the Pitot probe measurement, wind speed ($\textbf{V}_{w}$), airship groundspeed ($\textbf{V}_{g}$) and orientation ($\boldsymbol{\Phi}$).
	
	Let the airship motion be represented by its inertial velocity $\textbf{V}_{g}$. Similarly, the wind is described by an inertial velocity $\textbf{V}_w$. The airship relative air velocity is called airspeed ($\textbf{V}_a$) and it is given by:
	\begin{equation}
	\label{eq:vec_va}
	\textbf{V}_a = \textbf{V}_{g} - \textbf{V}_w\mathrm{,}
	\end{equation}
	where $\textbf{V}_w = [u_w\, v_w\, w_w]^T$ and $\textbf{V}_a = [u_a\, v_a\, w_a]^T$.
	
	The Euclidean norm of the airspeed is called \textit{true airspeed} ($V_t$) and it is given by:
	\begin{equation}
	\label{eq:vt}
	V_t = ||\textbf{V}_a||_2 = \sqrt{u_a^2 +v_a^2 + w_a^2} \mathrm{.}
	\end{equation}
	
\begin{figure}[htb]
		\centering
		\includegraphics{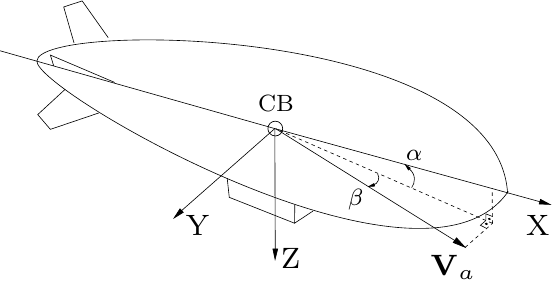}
		\caption{Sideslip angle ($\beta$) and angle of attack ($\alpha$).}
		\label{fig:fig_aa_bb}
	\end{figure}

	Other important definitions are the sideslip angle $\beta$ and angle of attack $\alpha$. The sideslip angle is a relative orientation between the vertical plane of the vehicle and the vector $\textbf{V}_a$. Moreover, the angle of attack $\alpha$ is given by the angle between the vector $\textbf{V}_a$ and the horizontal plane of the vehicle, as shown in Figure \ref{fig:fig_aa_bb}.
	
	Threfore, we can define $\beta$ and $\alpha$ by the following statements:
	\begin{equation}
	\label{eq:beta}
	\beta = \sin^{-1}\frac{v_a}{V_t}\mathrm{,}
	\end{equation}
	\begin{equation}
	\label{eq:alpha}
	\alpha = \tan^{-1}\frac{w_a}{u_a}\mathrm{.}
	\end{equation}
	
	Thus, an equivalent formula is given by:
	\begin{equation}
	w_a = u_a\frac{\sin \alpha}{\cos \alpha} \;\;\; \mathrm{and}\;\;\; v_a = V_t \sin \beta\mathrm{.}
	\end{equation}
	
	Finally, we obtain:
	
	\begin{equation}
	\label{eq:vt2_ua2}
	V_t  = \frac{u_a}{\cos \alpha \cos \beta}\mathrm{.}
	\end{equation}
	
	The Pitot tube is located in the airship nose, thus it will be measuring the longitudinal dynamic pressure $\Delta P$ in the body frame. Also, it has a correlation with the airspeed as shown below:
	\begin{equation}
	\label{eq:dp_ua}
	\Delta P  =\eta \big(u_a\big)^2\mathrm{,}
	\end{equation}
	where $\eta$ is the calibrating factor that is correlated with the air density and pitot efficiency. Now consider the following variable transformation:
	\begin{equation}
	\label{eq:v_pitot}
	V_{pitot} = \sqrt{\Delta P}\mathrm{,}
	\end{equation}
	Thus, $V_{pitot}$ is correlated with the true airspeed through the following statement:
	\begin{equation}
	\label{eq:vt_square}
	V_t  = \frac{V_{pitot}}{\sqrt{\eta }\cos \alpha\cos \beta}\mathrm{,}
	\end{equation}
	Because there are uncertainties in $\eta$ and the angles $\alpha$ and $\beta$ are unmeasurable with the actual available sensors, those values will be estimated together as a scale factor $c_f$ given by:
	\begin{equation}
	\label{eq:cf}
	c_f = \sqrt{\eta}\cos\alpha \cos\beta\mathrm{,}
	\end{equation}
	therefore \eqref{eq:vt_square} becomes:
	\begin{equation}
	V_t  = \frac{1}{c_f} V_{pitot}\mathrm{.}
	\end{equation}
	
	Now, consider that the rotation of vector $\textbf{V}_g$ from the body frame to the NED frame is given by $\textbf{V}_{NED} = [V_N~V_E~V_D]^T$. Also, consider that such rotation applied to vector $\textbf{V}_w$ is given by $\textbf{V}_{NED{w}}$. Thus, the airspeed in NED frame is given by:
	\begin{equation}
	\textbf{V}_{NEDa} = \textbf{V}_{NED} - \textbf{V}_{NEDw} = S_\Phi^{T} \textbf{V}_{a}\mathrm{.}
	\end{equation}
	
	It is known that a rotational operation does not change the vector module, thus the following statement is valid, considering the wind strictly horizontal:
	\begin{equation}
	\label{eq:z1}
	V_{pitot}^2 =  c_f^2 \big((V_N - V_{N_w})^2 +(V_E - V_{E_w})^2 + (V_D)^2\big)\mathrm{.}
	\end{equation}
	
	Supposing that, the airship starts from an initial condition where $\alpha$ and $\beta$ are negligible ($\alpha \approx \beta \approx 0$) we have $u_a = V_t$ and $v_a=w_a=0$. Therefore in the global frame we have:
	
	\begin{equation}
	V_{N_a} = u_a \cos \psi \cos \theta\mathrm{,}
	\end{equation}
	\begin{equation}
	V_{E_a} = u_a \sin \psi \cos \theta\mathrm{,}
	\end{equation}
	where $\psi$ and $\theta$ are the yaw and pitch angles, respectively. Since we have \eqref{eq:vec_va}, then:
	
	\begin{equation}
	\label{eq:z2}
	V_N = \frac{V_{pitot}}{c_f}  \cos \psi \cos \theta + V_{N_w}\mathrm{,}
	\end{equation}
	\begin{equation}
	\label{eq:z3}
	V_E = \frac{V_{pitot}}{c_f}   \sin \psi \cos \theta + V_{E_w}\mathrm{.}
	\end{equation}
	
	The values of $V_{pitot}$, $V_N$ and $V_E$ are measurable by the Pitot tube and GPS, therefore \eqref{eq:z1}, \eqref{eq:z2} and \eqref{eq:z3} can be used as observation equations, while $c_f$, $V_{N_w}$ and $V_{E_w}$ are estimated states in the EKF. Note that, the Euler angles ($\phi$, $\theta$ and $\psi$) can be measured by the IMU. 
	
	\section{Extended Kalman Filter}
	\label{sec:ekf_design}

		In this section we propose an EKF in order to estimate the incident wind in the airship body. Assuming that the wind is strictly horizontal the main goal is to obtain an estimation of the wind velocity in the horizontal plane (North-East) and the Pitot probe scale factor. One main benefit is expected from this method. The estimator presented in \cite{cho2011}, uses only one measurement update equation which is similar to \eqref{eq:z1}. By introducing \eqref{eq:z2} and \eqref{eq:z3} in the measurement update stage we expect to have a better performance than using only \eqref{eq:z1}.

		There is no given model to determine the wind behavior. Thus, we assume that the wind is constant with a Gaussian input with significant covariance. In addition, as long as we do not have a sideslip sensor, thus the uncertain and time varying factor $c_f$ will be estimated, which leads us to the following reduced system model:

		\begin{subequations}
			\begin{equation}
			\label{eq:speed_syn}
			\boldsymbol{\chi}_{k+1} = \textbf{F} \boldsymbol{\chi}_k + \mathfrak{\boldsymbol{\nu}}_k\mathrm{,}
			\end{equation}
			\begin{equation}
			\label{eq:speed_syn_out}
			\textbf{z}_k = \textbf{h}(\boldsymbol{\chi}_k) + \boldsymbol{\upsilon}_k
			\end{equation}
		\end{subequations}
		where: $\boldsymbol{\chi}_k = [V_{N_{wk}} ~V_{E_{wk}}~c_{fk}]^T$ is the state vector in the instant $t = k t_s$; $\textbf{z}_k=[V_{pitot_k}^2~V_{N_k}~V_{E_k}]^T$ is the system output in the instant $t = k t_s$; $t_s$ is the sample time in seconds; $\textbf{h}(\boldsymbol{\chi}_k)$ is the output function, which can be computed through $\eqref{eq:z1}$, $\eqref{eq:z2}$ and $\eqref{eq:z3}$;  
		\begin{equation*}
		\textbf{F} = \begin{bmatrix}
		1 & 0 & 0\\
		0 & 1 & 0\\
		0 & 0 & 1
		\end{bmatrix}\text{; } 
		\end{equation*}
		$\mathfrak{\boldsymbol{\nu}}_k \sim \textbf{N}(0,\textbf{Q})$ is the process noise with Gaussian distribution and covariance $\textbf{Q}$; and, finally, $ \mathfrak{\boldsymbol{\upsilon}}_k \sim \textbf{N}(0,\textbf{R})$ is the measurement noise also with Gaussian distribution and covariance $\textbf{R}$.

	Given the model described in \eqref{eq:speed_syn}, we can update the state and covariance matrix ($\textbf{P}$) as follows:
	\begin{equation}
	\boldsymbol{\chi}_{k|k-1} = \textbf{F} \boldsymbol{\chi}_{k-1}\mathrm{,}
	\end{equation}
	\begin{equation}
	\textbf{P}_{k|k-1} = \textbf{F} \textbf{P}_{k-1} \textbf{F}^T + \textbf{Q}\mathrm{.}
	\end{equation}

		For the accomplishment of the EKF final stage we obtain the Jacobian matrix of $\textbf{h}(\boldsymbol{\chi}_k)$ evaluated in the measured values of $V_{pitot}$, $\textbf{V}_{NED}$, $\boldsymbol{\Phi}$ and $\boldsymbol{\chi}_{k|k-1}$ which is given by:

		\begin{equation*}\textbf{H}_k =\left[ \frac{\partial \textbf{h}(\boldsymbol{\chi})}{\partial V_{N_{w}}}, ~\frac{\partial \textbf{h}(\boldsymbol{\chi})}{\partial V_{E_{w}}}, ~\frac{\partial \textbf{h}(\boldsymbol{\chi})}{\partial c_f}\right]\Bigg|_{\boldsymbol{\chi}_{k|k-1}},
		\end{equation*} where:
	
	\begin{equation*}
	\frac{\partial \textbf{h}(\boldsymbol{\chi})}{\partial V_{N_{w}}} = [-2\hat{c}_f^2\hat{V}_{N_w}(V_N - \hat{V}_{N_w})~1~0]^T\mathrm{,}
	\end{equation*}
	\begin{equation*}
	\frac{\partial \textbf{h}(\boldsymbol{\chi})}{\partial V_{E_{w}}} = [-2\hat{c}_f^2\hat{V}_{E_w}(V_E - \hat{V}_{E_w})~0~1]^T ~~\text{and}
	\end{equation*}
	\begin{equation*}
	\frac{\partial \textbf{h}(\boldsymbol{\chi})}{\partial c_f} = \begin{bmatrix}
	2\hat{c}_f\big((V_N - \hat{V}_{N_w})^2 + (V_E - \hat{V}_{E_w})^2 + (V_D)^2\big)\\
	-\frac{V_{pitot}}{\hat{c}_f^2}  \cos \psi \cos \theta\\
	-\frac{V_{pitot}}{\hat{c}_f^2}  \sin \psi \cos \theta
	\end{bmatrix}\mathrm{.}
	\end{equation*}
	
	Finally, the standard algorithm of EKF can be applied as follows:
	\begin{equation*}
	\tilde{\textbf{y}}_k = \textbf{z}_k - \textbf{h}(\boldsymbol{\chi}_{k|k-1})\mathrm{,}
	\end{equation*}
	\begin{equation*}
	\textbf{C}_k = \textbf{H}_k \textbf{P}_{k|k-1}\textbf{H}_k^T+\textbf{R}\mathrm{,}
	\end{equation*}
	\begin{equation*}
	\textbf{K}_k = \textbf{P}_{k|k-1}\textbf{H}_k^T \textbf{C}_{k}^{-1}\mathrm{,}
	\end{equation*}
	\begin{equation*}
	\boldsymbol{\chi}_k=\boldsymbol{\chi}_{k|k-1}+\textbf{K}_k\tilde{\textbf{y}}_k\mathrm{,}
	\end{equation*}
	\begin{equation*}
	\textbf{P}_k= (\textbf{I}-\textbf{K}_k \textbf{H}_k)\textbf{P}_{k|k-1}\mathrm{,}
	\end{equation*}
	where $\textbf{P}$ is the covariance matrix, $\textbf{C}$ is the covariance error, $\tilde{\textbf{y}}$ is the measurement error, $\textbf{K}$ is the Kalman gain and $\textbf{I}$ is the identity matrix with appropriate dimensions.
	
	\section{Neural Network}
	\label{sec:nn_design}

		This  section  proposes a Neural Network (NN) to estimate the wind speed in NED frame and the Pitot probe scale factor $c_f$. The main benefit expected from this method is the NN intrinsic property of create a mapping from input parameters to the output. Because  the  NN  is  trained  by  measured  data,  it is able to detect abrupt variation in the wind velocity. However, it is also expected to be very sensitive to measurement errors.

		As  the  analytical  model \eqref{eq:speed_syn}--\eqref{eq:speed_syn_out} shows, the model output is composed by \eqref{eq:z1}, \eqref{eq:z2} and \eqref{eq:z3}, which are nonlinear equations in the model states, vehicle velocity, orientation and Pitot pressure. In order to avoid complex nonlinearities, the measured data is remapped into 8 inputs given by:
	
	\begin{equation*}
	\textbf{z}_{nn} = \begin{bmatrix}
	V_{pitot}^2\\
	V_D^2\\
	V_N\\
	V_E\\
	V_E^2\\
	V_N^2\\
	V_{pitot} \cos \psi \cos \theta\\
	V_{pitot} \sin \psi \cos \theta
	\end{bmatrix}\mathrm{.}
	\end{equation*}
	
	Meanwhile the output vector $\boldsymbol{\chi}_{nn}$ of the NN are the estimated wind velocities in the horizontal plane and the scale factor $c_f$ as shown below:
	\begin{equation*}
	\boldsymbol{\chi}_{nn} = \begin{bmatrix}
	V_{N_w}\\
	V_{E_w}\\
	c_f
	\end{bmatrix}\mathrm{.}
	\end{equation*}
	
\begin{figure}[htb]
		\centering
		\includegraphics[scale=0.95]{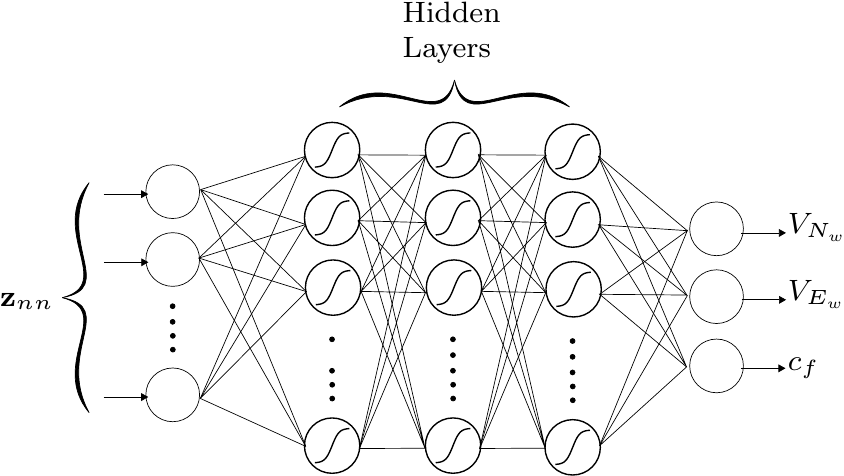}
		\caption{Neural Network flow chart.}
		\label{fig:generic_nn}
	\end{figure}

	The NN was designed in the MATLAB Neural Network Toolbox$^{TM}$. It is a three-layer fitting NN, which has three nonlinear hidden layers containing 24 neurons each and three linear outputs. The activation function of the nonlinear neurons are sigmoidal. The resulting flow chart is shown in Figure \ref{fig:generic_nn}.
	
	
\begin{figure}[htb]
		\centering
		\includegraphics[scale=0.64]{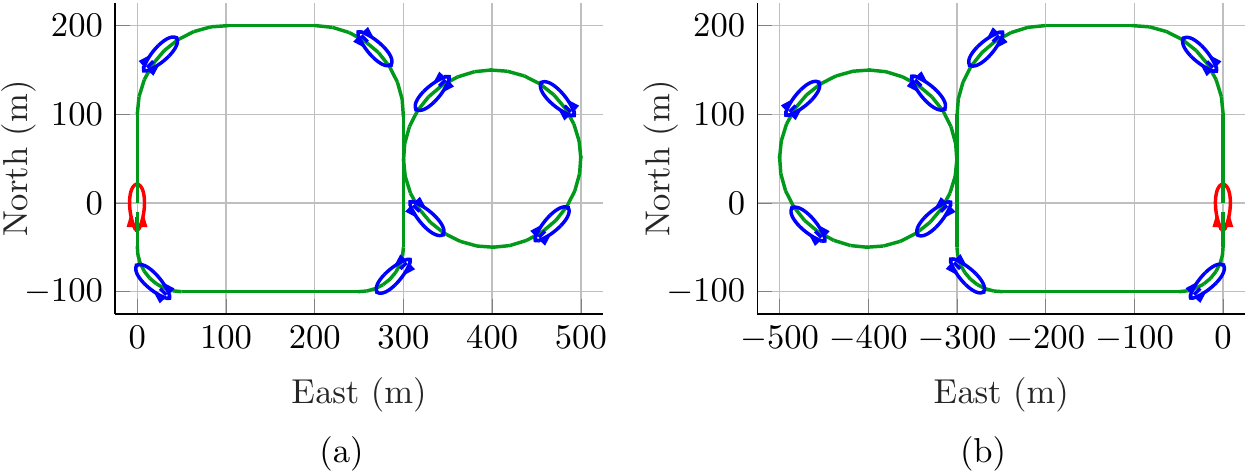}
		\caption{Training missions: (a) first path and (b) second path.}
		\label{fig:trainning_traj}
	\end{figure}

\begin{figure*}
		\centering
		\includegraphics[scale=0.67]{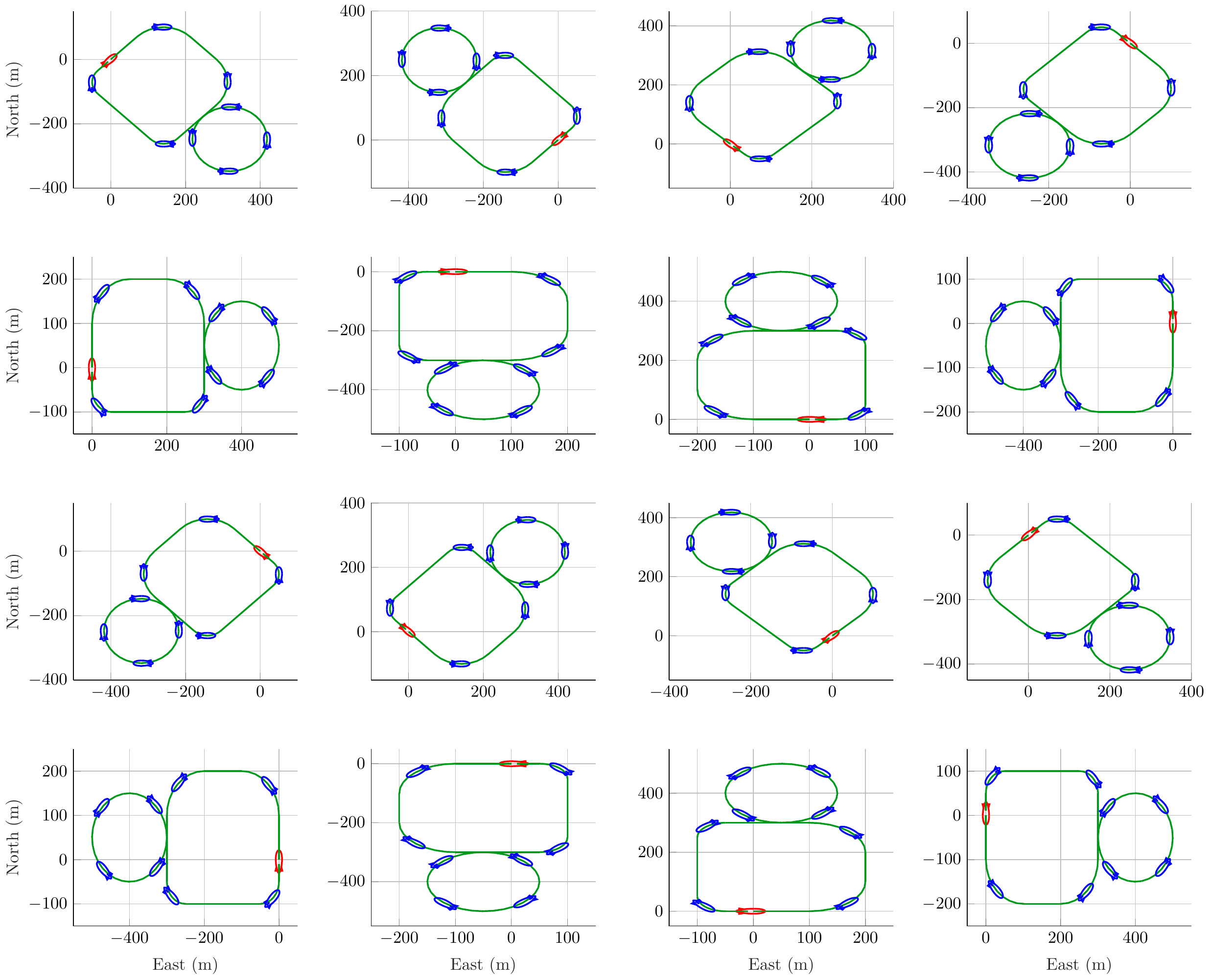}
		\caption{Simulation trajectories used for the NN training task}
		\label{fig:training_traj_all}
	\end{figure*}

		The training dataset is composed by simulations in variations of the two trajectories shown in Figure \ref{fig:trainning_traj}. The trajectories (a) and (b) are rotated of \{0, 45, 90, 135,  180, 225, 270, 315\} degrees around origin generating 16 scenarios (see Figure \ref{fig:training_traj_all}) in which the airship performs curves and straight lines in different directions. For each scenario were performed simulations with wind speed at $|\vec{V}_w|=$ \{0, 1, 2, 3, 4, 5\}m/s and heading $\phi_w=$\{0, 22.5, 45, 67.5, 90, 112.5, 135, 157.5, 180, 202.5, 225, 247.5, 270, 292.5, 315, 337.5\} degrees, where $\phi_w = \tan^{-1}\big(\frac{V_{E_w}}{V_{N_w}}\big)$. Hence, a total of 1281 simulations were performed. In all simulations the airship performs a typical cruise flight at 7m/s airspeed and constant altitude of 50 meters. 
	
		The dataset stored has approximately 375 MB. Training phase and further evaluations were carried out  on  a  desktop  architecture  which  features  a  four-core 4.00 GHz Intel Core i7-6700K Processor, NVIDIA GeForce GTX 950, 32GB of RAM and Ubuntu 16 LTSOS. The system took 22 minutes and 17 seconds to train the afore-mentioned dataset.

		Training was achieved using Matlab Neural Network Toolbox. The Scaled Conjugate Gradient (SCG) algorithm is used with 5000 epochs of training iterations using 70\% of the collected data randomly taken as the training set, 15\% used for validation set and 15\% as the test set. The performance evaluation is made by Mean Squared Error (MSE).

\begin{figure}[htb]
		\centering
		\includegraphics[scale=0.7]{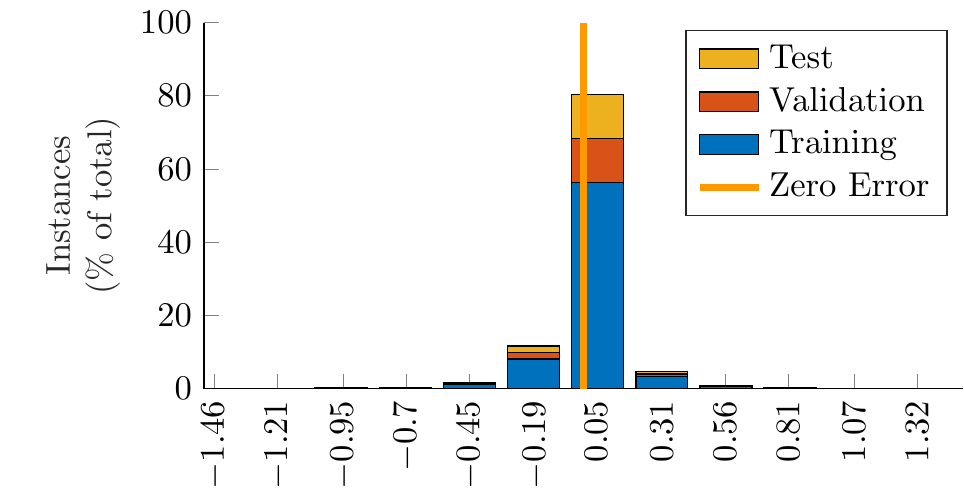}
		\caption{Error histogram of the NN dataset}
		\label{fig:wind_estimation_error_hist}
	\end{figure}

		The residual error of the trained NN is shown in Figure \ref{fig:wind_estimation_error_hist}. Note that, about 80\% of the total data are distributed around zero, in which 60\% were taken as training data, 10\% as validation data and about 10\% as test data.

		The correlation coefficient (R-value), which is a linear regression between the NN predicted values and the targets, and best MSE are shown in Table \ref{tab:correlation}. Values of R closer to 1 indicate better agreement between targets and predicted values. Note that both R-value and MSE are close for Training, Validation and Test dataset, which indicates that the NN is not overfitting the data.
	
	\begin{table}[htb]
		\caption{Correlation coefficient R-value and MSE.}
		\label{tab:correlation}
		\centering
		\begin{tabular}{c c c}
			\hline \hline
			Dataset & R-value  & MSE \\
			\hline \hline
			Training & 0.99731 & 0.0206\\ 
			\hline
			Validation & 0.99732 & 0.0205\\ 
			\hline
			Test&  0.99720 & 0.0205\\ 
			\hline
			Total & 0.99729 & 0.0205\\
			\hline \hline
		\end{tabular}
	\end{table} 
	
\begin{figure}[htb]
		\centering
		\includegraphics{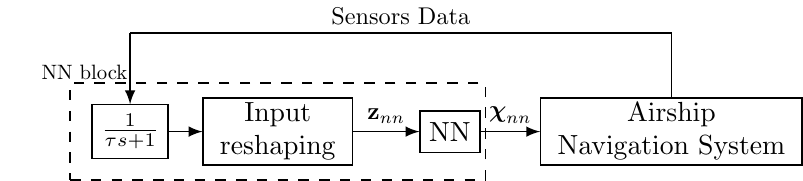}
		\caption{Resulting block diagram of the NN approach}
		\label{fig:blk_diag_nn}
	\end{figure}

		In order to avoid high frequency oscillations in the estimation, a low-pass filter  was introduced for filtering the input values $\textbf{V}_g$, $\boldsymbol{\Phi}$ and $V_{pitot}$. The low-pass filter can be expressed by the following Laplace representation:
		\begin{equation}
		\frac{Y(s)}{Z(s)} = \frac{1}{\tau s + 1}
		\end{equation}
		where $\tau$ is the time constant (here we use $\tau=1.5$ seconds), $Z(s)$ is the Laplace transform of the input and $Y(s)$ is the Laplace transform of the filter output. After filtering the measured sensor data, the input vector $\textbf{z}_{nn}$ is computed and passed by the trained NN as shown in the block diagram from Figure \ref{fig:blk_diag_nn}. Finally, the estimated wind velocity $\boldsymbol{\chi}_{nn}$ is given to the airship navigation system.

	\section{Hybrid estimator}
	\label{sec:hybrid_design}
	
\begin{figure}[htb]
		\centering
		\includegraphics{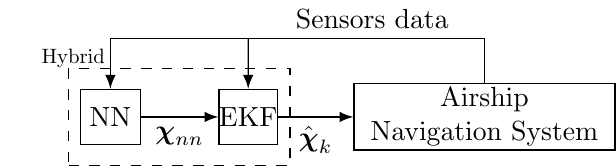}
		\caption{Hybrid estimator with cascaded form.}
		\label{fig:blk_diag_hybrid}
	\end{figure}

	Here we propose a hybrid estimator that performs a fusion between both estimators, namely: from the EKF designed in Sect. \ref{sec:ekf_design} and NN designed in Sect. \ref{sec:nn_design}. This fusion is performed by changing the measure update stage of the EKF approach. The NN output $\boldsymbol{\chi}_{nn}$ is added to the measurement vector of the EKF as a redundant measure. Thus, resulting in the new measurement vector $\textbf{z}_{h_k}$, updating function $\textbf{h}_{h_k}(\boldsymbol{\chi}_k)$ and its respective Jacobian $\textbf{H}_{h_k}$ shown below:
	\begin{equation*}
	\textbf{z}_{h_k}=\begin{bmatrix}\textbf{z}_k\\\boldsymbol{\chi}_{nn}\end{bmatrix}\mathrm{,}~\textbf{h}_{h_k}(\boldsymbol{\chi}_k) =\begin{bmatrix}
	\textbf{h}(\boldsymbol{\chi_k})\\\boldsymbol{\chi}_{k}\end{bmatrix}~\mathrm{and} ~\textbf{H}_{h_k}=\begin{bmatrix}\textbf{H}\\\textbf{I}_3\end{bmatrix}\mathrm{,}
	\end{equation*}
	where $\textbf{I}_3$ is the identity matrix of third order. Then the EKF standard algorithm is used by updating the dimensions of the matrices $\textbf{C}_k$, $\textbf{K}_k$ and $\textbf{R}$. The resulting estimator has a cascaded form as illustrated in Figure \ref{fig:blk_diag_hybrid}.
	
	It is important to highlight that the NN used here is the same NN previously designed and trained in Sect. \ref{sec:nn_design}.
	
	\section{Simulation results}
	\label{sec:results}
		In this section two simulations are performed in order to evaluate all the three approaches presented before and establish a comparison with the traditional model-based approach proposed by Cho et. al. \cite{cho2011}.

\begin{figure}[htb]
		\centering
		\includegraphics{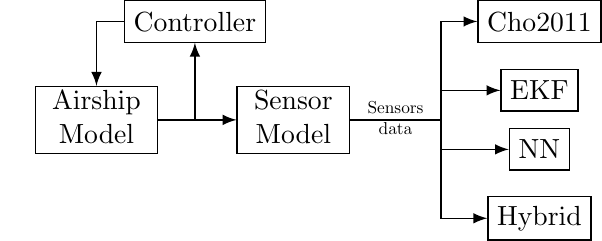}
		\caption{Block diagram of simulation}
		\label{fig:blk_diag_simulation}
	\end{figure}

\begin{figure}[htb]
		\centering
		\includegraphics{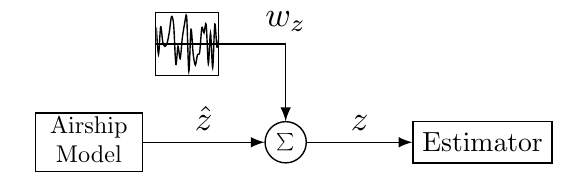}
		\caption{Block diagram of a generic sensor modeling.}
		\label{fig:generic_sensor}
	\end{figure}

	During simulation the airship is well controlled with ideal feedback, meanwhile the estimators evaluated are receiving noisy data from the modeled sensors as shown in the block diagram from Figure \ref{fig:blk_diag_simulation}. 	The sensors are modeled in the simulation environment with sample frequency as specified in Table \ref{tab:sample_freq}. Also, each sensor has a generic modeling including a Gaussian noise as specified by the manufacturer in Table \ref{tab:sensor_noise}. The generic sensor modeling is shown in Figure \ref{fig:generic_sensor}, where $w_z$ is a Gaussian noise,  $\hat{z}$ is the true simulated value and $z$ is the sensor output. All estimators use the same sample frequency of 16Hz ($t_s=0.0625$ seconds). 


	\begin{table}[htb]
		\caption{Sensor noise standard deviation}
		\label{tab:sensor_noise}
		\centering
		\begin{tabular}{c c}
			\hline \hline
			Data  & $\sigma(w_y)$  \\
			\hline \hline
			$\phi, \theta$ (rad) & 5.2$\cdot 10^{-3}$  \\ 
			\hline 
			$\psi$ (rad) & 0.1 \\
			\hline 
			$\textbf{V}_g$ (m/s) & 0.4 \\ 
			\hline 
			$V_{pitot}$ (m/s) & 6.04$\cdot 10^{-4}$ \\ 
			\hline \hline
		\end{tabular}
	\end{table}
	\begin{table}[htb]
		\caption{Sample frequency specification}
		\label{tab:sample_freq}
		\centering
		\begin{tabular}{c | c c }
			\hline \hline
			Sensor  & Frequency & Sampled data\\
			\hline \hline
			IMU & 100 Hz & $\boldsymbol{\Phi}$ \\ 
			\hline 
			GPS & 4 Hz & $\textbf{V}_g$\\ 
			\hline 
			Pitot tube & 18 Hz & $V_{pitot}$\\ 
			\hline\hline
		\end{tabular}
	\end{table}

		An online repository\footnote{https://github.com/leve-fem/airship\_estimator} is available containing all approaches presented here. The algorithms are implemented in C/C++ and Python inside the Robot Operating System (ROS) \cite{ros}. Also, in this same repository a link to the dataset used for the NN training task and the simulations performed are available for future researches. 
	
	\subsection{First scenario}
		The first scenario considers wind with absolute value $|\vec{V}_w|=2m/s$ and heading $\psi_w=\frac{\pi}{2}$ rad (blowing from East to West). Then in the instant $t=160s$ the wind is intensified to  $|\vec{V}_w|=3m/s$ and its heading is changed to $\psi_w= \pi$ (blowing from South to North). 

\begin{figure}[htb]
		\centering
		\includegraphics[angle=0, scale=0.75]{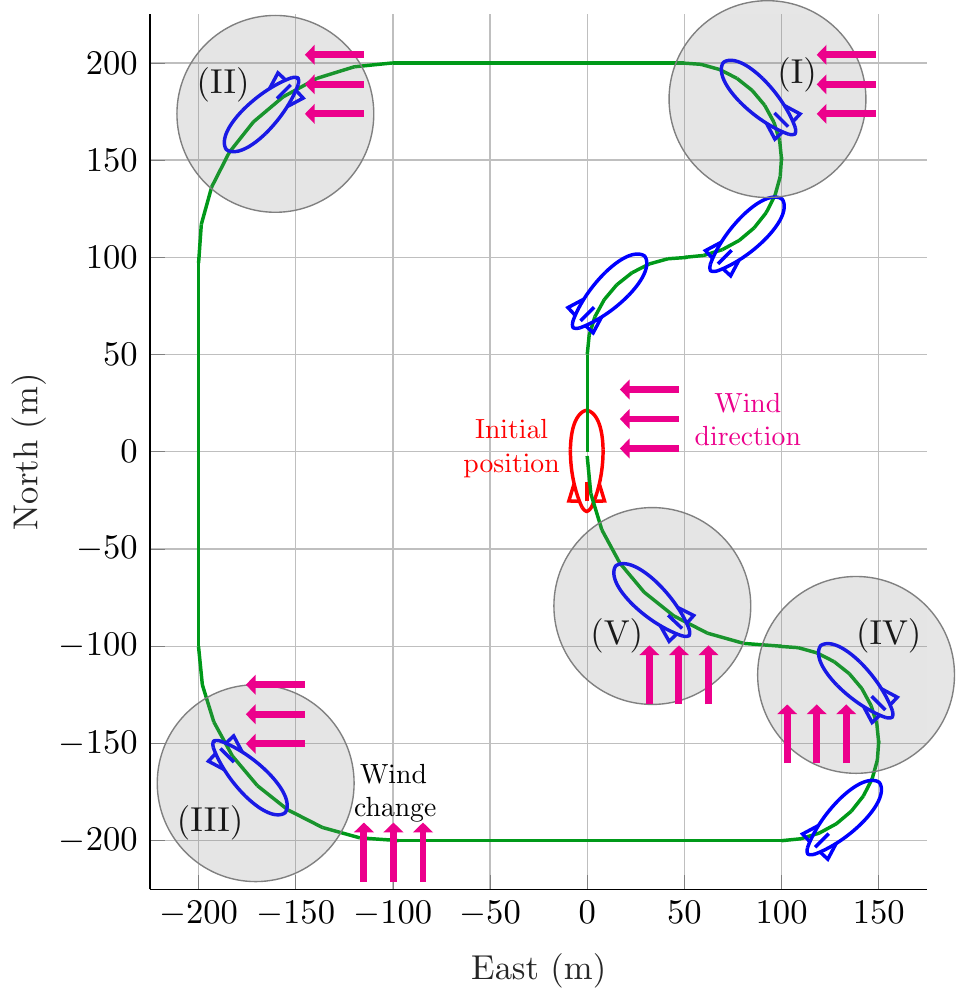}
		\caption{Simulation trajectory.}
		\label{fig:validating_traj}
	\end{figure}

\begin{figure}[htb]
		\centering
		\includegraphics{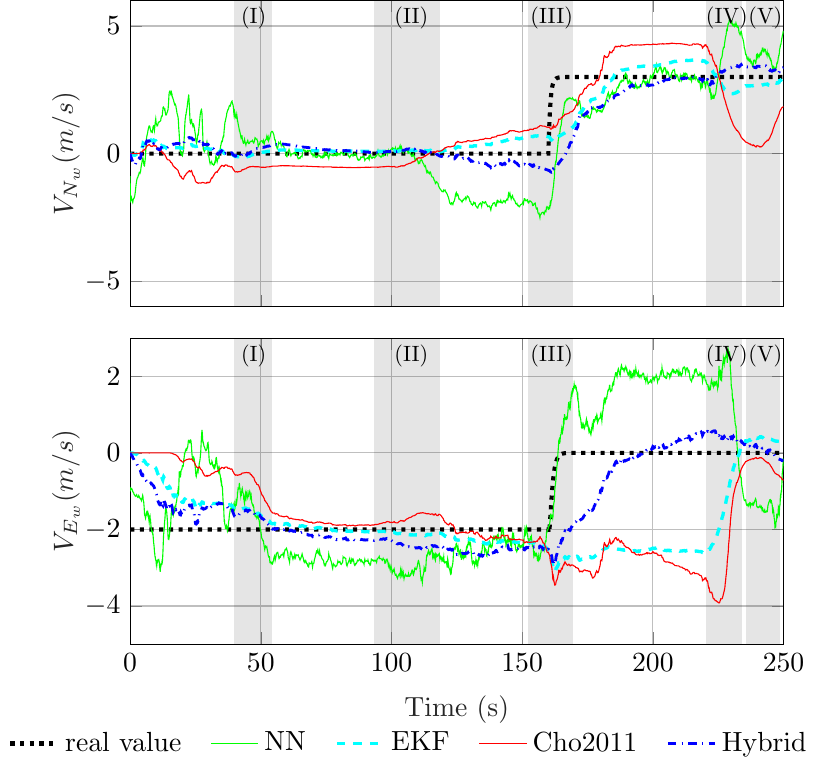}
		\caption{First scenario of simulation in the airship nonlinear model with realistic sensor noise: wind velocity estimation in North-East frame.}
		\label{fig:wind_estimation_velocities}
	\end{figure}

		The airship is controlled to follow the trajectory shown in Figure \ref{fig:validating_traj}. Five instants are highlighted with gray background in order to establish further comparisons with the results in Figure \ref{fig:wind_estimation_velocities}. Moreover, results using the estimator proposed by \cite{cho2011} were introduced as ``Cho2011'' in order to establish a comparison. The covariance matrices used in the Model-based approaches can be found in Appendix.

	In Figure \ref{fig:wind_estimation_velocities} is possible to note that, the ``Cho2011'' estimator has to acquire information about the motion in all directions before it converges to the correct values. After the instant (I), all estimators converges to values within of an acceptable error. It is important to note that, the NN has some high frequency oscillations at the trajectory curves, which deteriorate the performance. When the airship is following a straight line the designed NN has a good estimation, however sometimes with a bias from the real value.
	
	Also in Figure \ref{fig:wind_estimation_velocities}, we can note that when the wind velocity has a significant variation in the instant (III), the two Model-based approaches (``EKF'' and ``Cho2011'') do not converge immediately because both depend on information (given by the Pitot tube) about the other directions to converge to the correct wind velocity. Meanwhile the NN clearly has an instantly reaction to these variations. Even though the NN converges for a biased value, such information was sufficient to correct the estimation of the Hybrid approach before the instants (IV) and (V). In these final instants, the  Model-based estimators finally converges for values within a range of acceptable error.
	
	In Table \ref{tab:rms_validating} are shown the RMS values of the estimation errors $\tilde{V}_{N_w} = \hat{V}_{N_w} - V_{N_w} $ and $\tilde{V}_{E_w} = \hat{V}_{E_w} - V_{E_w}$ and the percentage in relation to the approach presented by Cho et. al.\cite{cho2011}. By the RMS values we can observe that the NN had a better estimation of $V_{E_w}$ in comparison to the Model-based approaches. However, for the component $V_{N_w}$, the Model-based approaches presented a better performance. Meanwhile the ``Hybrid'' which has the information of both approaches had the best performance in the estimation of $V_{E_w}$ and acceptable estimation in the $V_{N_w}$ component.

	\begin{table}[htb]

		\caption{RMS value of the estimation error.}
		\label{tab:rms_validating}
		\begin{tabular}{lll}
			\hline\noalign{\smallskip}
			& $\tilde{V}_{N_w}$ (m/s) & $\tilde{V}_{E_w}$ (m/s) \\
			\noalign{\smallskip}\hline\noalign{\smallskip}
			Cho2011 & 1.01 & 1.74\\
			EKF & 0.58 \textcolor{blue}{(-42.6\%)} & 1.42 \textcolor{blue}{(-18.3\%)}\\
			NN & 1.19 \textcolor{red}{(+17.8\%)} & 1.25 \textcolor{blue}{(-27.0\%)}\\
			Hybrid & 0.74 \textcolor{blue}{(-26.7\%)} & 0.71 \textcolor{blue}{(-59.2\%)}\\
			\noalign{\smallskip}\hline
		\end{tabular}
	\end{table}
	
\begin{figure}[htb]
		\centering
		\includegraphics[scale=0.67]{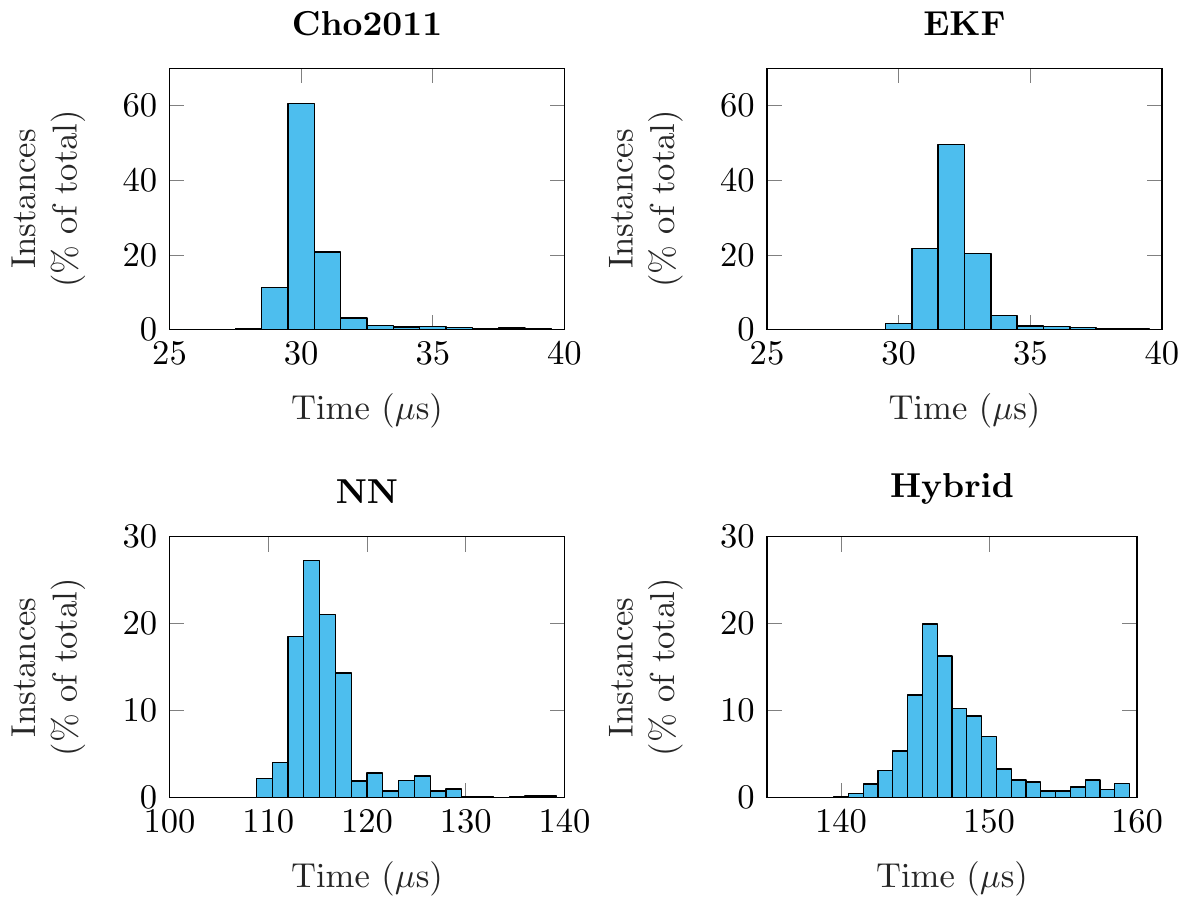}
		\caption{Computational time histogram}
		\label{fig:wind_estimation_computational_cost}
	\end{figure}

		Figure \ref{fig:wind_estimation_computational_cost} shows the histograms of computational time necessary for each method. Although NN and Hybrid approaches have higher computational time (100 $\sim$ 160 microseconds), they are able of running at almost 1000Hz. Meanwhile, both Model-based approaches (EKF and Cho2011) are able of running at more than 1000Hz with a computational time between 25 and 35 microseconds. Note that the proposed approach by Cho et. al. \cite{cho2011} has lowest computational cost, since the EKF proposed here has two additional measurement update equations.

	\subsection{Second scenario}
	
\begin{figure}[htb]
		\centering
		\includegraphics{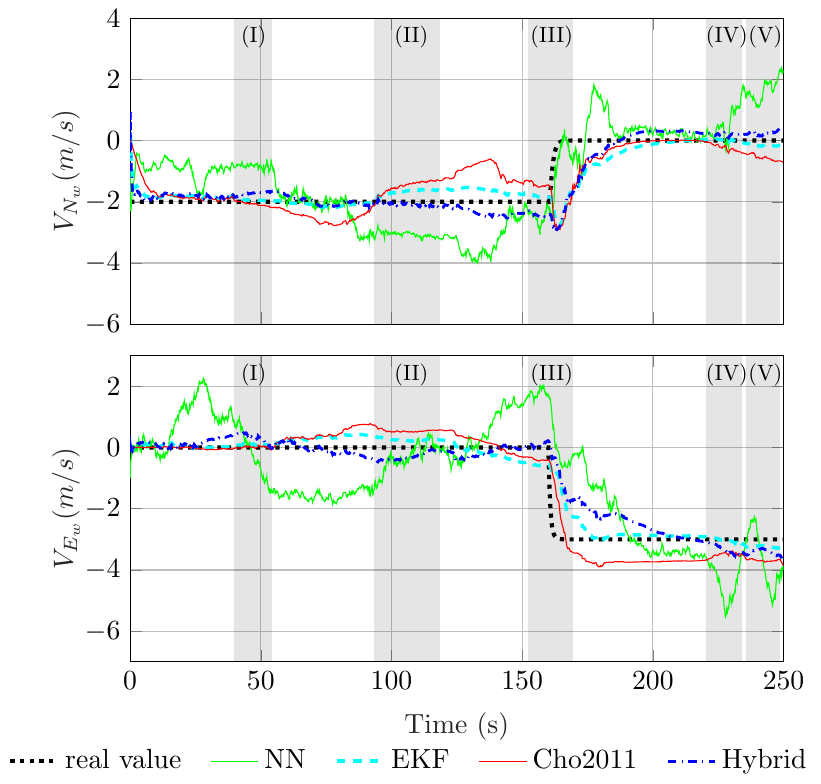}
		\caption{Second scenario of simulation in the airship nonlinear model with realistic sensor noise: wind velocity estimation in North-East frame.}
		\label{fig:wind_estimation_velocities_12}
	\end{figure}

		In a second simulation in the same trajectory of the previous simulation, it is considered a different variation in the wind velocity. Initially the wind has absolute value $|\vec{V}_w|=2m/s$ and heading $\psi_w=0$ (blowing from North to South), then in the instant $t=160s$ the wind is intensified to  $|\vec{V}_w|=3m/s$ and its heading is changed to $\psi_w= \frac{\pi}{2}$rad (blowing from East to West). The results are shown in Figure \ref{fig:wind_estimation_velocities_12} and resultant RMS error is shown in Tab. \ref{tab:rms_validating2}. 

	\begin{table}[htb]	
		\centering
		\caption{RMS value of the estimation error in second simulation.}
		\label{tab:rms_validating2}
		\begin{tabular}{lcc}
			\hline\noalign{\smallskip}
			& $\tilde{V}_{N_w}$ (m/s)& $\tilde{V}_{E_w}$ (m/s) \\
			\noalign{\smallskip}\hline\noalign{\smallskip}
			Cho2011 & 0.71 &  0.52\\
			EKF & 0.52 \textcolor{blue}{(-26.5\%)}  &  0.38 \textcolor{blue}{(-26.9\%)}\\
			NN & 1.01 \textcolor{red}{(+42.2\%)}  &  1.21 \textcolor{red}{(+132.7\%)}\\
			Hybrid & 0.46 \textcolor{blue}{(-35.2\%)}  &  0.52 \textcolor{black}{(0\%)}\\
			\noalign{\smallskip}\hline
		\end{tabular}
	\end{table}

		In this second simulation, the airship starts with the nose against the wind, thus a few seconds of simulation are sufficient to the Model-based approaches estimate correctly the wind velocity components. Note that as the airship do curves in the scenario in instants (I) and (II), NN and Cho2011 approaches presents some estimation errors, while the EKF and Hybrid maintain the estimation near the true value. The improvement made in the model-based approaches by including two additional measurement update equations avoids high oscillations. 
	
		In the overall simulation, Model-based approaches had better performance. It occurs because the observability is not degraded during the wind transition at instant $t=160$ seconds. After the instant (III), the airship is traveling to the East, thus the wind is against its nose. Therefore, the Pitot tube is able to capture the relative speed perfectly. By table \ref{tab:rms_validating2} we can also note that the EKF presented similar estimation performance for both wind components ($V_{N_w}$ and $V_{E_w}$), differently from the previous simulation. Clearly, in this scenario, the NN is not too advantageous. It presented the higher RMS errors.

	\section{Conclusions}
	\label{sec:conclusion}
	
	In this paper we presented Model-based and Data-driven approaches for estimation of wind velocity for a robotic airship. The Model-based approach uses only kinematic equations of motion of the airship for the design of an EKF. The Data-driven proposed approach is composed by a NN trained with a dataset containing 1281 simulations in different conditions. Also, a novel Hybrid approach is proposed, by performing a fusion between the designed Model-based and Data-driven approaches with a cascaded structure.

		A high-fidelity and realistic airship dynamic model for Matlab/Simulink  platform was used in the simulations. Two simulation scenarios were presented. In the first, the wind variation is not detected properly by the model-based approaches, thus the Data-driven is advantageous in this scenario. In the second, the Model-based are more precise in the estimation, and the Data-driven performance was the same or worse.  

		The performance results obtained showed that the proposed EKF has a slightly better performance in comparison to the strategy found in the literature. It occurs because of the two additional measurement update equations that we introduced. Although, the performance was improved, EKF still have the same drawbacks when in situations of low observability. On the other hand, the NN presents higher sensibility to wind variations in such situations, however with biased estimations and high frequency oscillations due to sensor noise and non-trained situations. The NN approach is independent of any mathematical model, however, requires a sufficient and representative database.

		The Hybrid approach maintains satisfactory performance in the two simulation scenarios presented. This approach combines the advantages of both estimators (model-based and data-driven) and mitigates the drawbacks, once it has the information about the model and about previous missions (training dataset). However, it requires a representative dataset as well as a mathematical modeling.

	As a general result, we can conclude that the cooperation between both approaches (Model-based and Data-driven) can be highly effective for solving estimation problems with observation deficiency.  For the specific problem of wind estimation we obtained satisfactory results with the Hybrid approach. However, since  the  proposed  methods  are  only  analyzed  theoretically and validated via simulation, an actual benchmark or field test is needed in the subsequent work to verify  the  proposed  approaches. Future efforts will be made to validate these results outside of a simulation environment.

	\section*{Appendix}

	The covariance matrices used for each Model-based approach are shown below:
	
	\begin{flalign*}
	&\text{Cho2011}:\\
	\textbf{R} &= \begin{bmatrix}
	163.84
	\end{bmatrix}\text{, }\\
	\textbf{Q}&= \mathrm{diag}\big(\begin{bmatrix}
	10^{-3}  & 10^{-4} &  5 \cdot 10^{-6}
	\end{bmatrix}\big)\mathrm{.}\\
	&\hfill\\
	&\text{EKF}: \\
	\textbf{R} &=  \mathrm{diag}\big(\begin{bmatrix}
	40.96 & 40.96  &  40.96
	\end{bmatrix}\big)\text{, }\\
	\textbf{Q}&= \mathrm{diag}\big(\begin{bmatrix}
	10^{-4} & 10^{-4} & ~5 \cdot 10^{-7}
	\end{bmatrix}\big)\mathrm{.}\\
	&\hfill\\
	&\text{Hybrid}: \\
	\textbf{R} &=  \mathrm{diag}\big(\begin{bmatrix}
	10.24 & 10.24 &  10.24 &  10.24 &  10.24 &  10.24 
	\end{bmatrix}\big)\text{, }\\
	\textbf{Q}&=\mathrm{diag}\big(\begin{bmatrix}
	10^{-4}  & 10^{-4}  &  ~5 \cdot 10^{-7}\\
	\end{bmatrix}\big)\mathrm{.}\\
	\end{flalign*}
	
	\begin{acknowledgements}
		The present work was sponsored by the Brazilian agencies CNPq and FAPESP, through Projects DRONI (CNPq 402112/2013-0), INCT-SAC (CNPq 465755 /2014-3; FAPESP 2014/50851-0) and scholarships (CNPq 305 600/2017-6; FAPESP BEP 2017/11423-0). Also, this work was supported by Funda\c c\~ao para a Ci\^encia e a Tecnologia (FCT), through IDMEC, under LAETA, project UID/EMS/ 50022/2013 (Portugal). Moreover, the authors are grateful to Erasmus Mundus SMART$^2$ for the financial support through project reference 552042-EM-1-2014-1-FR-ERA MUNDUS-EMA2. Furthermore, the authors appreciate all the support given by members of the Advanced Computing, Control \& Embedded Systems Laboratory (ACCES-Lab) and Laboratory of Study in Exterior Vehicles (LEVE) from University of Campinas. 
	\end{acknowledgements}
	
	%
	%

	\bibliographystyle{spmpsci}      

	
\end{document}